\documentclass[aps,prl,twocolumn,showpacs,superscriptaddress]{revtex4-1}  
\usepackage{graphicx}  
\usepackage{dcolumn}   
\usepackage{bm}        
\usepackage{amssymb}   
\usepackage{natbib}
\usepackage{color}

\begin{document}

\title{New dynamical scaling universality for quantum networks across adiabatic quantum phase transitions}

\author{O.~L. Acevedo}
\author{L. Quiroga}
\author{F.~J. Rodr\'{i}guez}
\affiliation{Departamento de F\'{i}sica, Universidad de los Andes, A.A. 4976, Bogot\'{a}, Colombia}
\author{N.~F. Johnson}
\affiliation{Department of Physics, University of Miami, Coral Gables, Miami, FL 33124, USA}

\begin{abstract}
We reveal universal dynamical scaling behavior across adiabatic quantum phase transitions (QPTs) in networks ranging from traditional spatial systems (Ising model) to fully connected ones (Dicke and Lipkin-Meshkov-Glick models). Our findings, which lie beyond traditional critical exponent analysis and adiabatic perturbation  approximations, are applicable even where excitations have not yet stabilized and hence provide a time-resolved understanding of QPTs encompassing a wide range of adiabatic regimes. We show explicitly that even though two systems may traditionally belong to the same universality class, they can have very different adiabatic evolutions. This implies more stringent conditions need to be imposed than at present, both for quantum simulations where one system is used to simulate the other,  and for adiabatic quantum computing schemes.\end{abstract}


\maketitle

Scaling is ubiquitous in nature, with critical exponents being used to characterize universal phase transition phenomena in both equilibrium and non-equilibrium systems \cite{Stanley}.
Scaling functions go beyond critical exponents by incorporating richer information about the dynamics of the underlying many-body system, including finite-size effects, and hence extending the range of validity over which theoretical predictions can describe empirical  data \cite{BeyondCrit}.
In the field of critical phenomena, much attention has recently been directed to adiabatic Quantum Phase Transitions (QPTs). In addition to their fundamental role as zero-temperature many-body quantum phenomena \cite{Sachdev}, QPTs represent a key ingredient of current quantum computation schemes \cite{AdQuCom1,*AdQuCom,Lidar1,*Lidar2}.

Recent studies show that as a QPT phase boundary is crossed slowly in models with finitely-connected lattices \cite{Dziarmaga,DziZurPRL,*DziZur,Cuchietti,Viola,TutDynAd,Huse}, the short-range interaction allows a correlation length to be defined and hence scaling to be examined through the Kibble-Zurek Mechanism (KZM) \cite{Zurek2,*Zurek,Dziarmaga}. However, an implicit limiting assumption of the KZM is that adiabatic evolution holds except for a small threshold around the critical point, during which spatial defects in the order-parameter are created and power-law relations emerge, defined by critical exponents. For slower quench rates, Adiabatic Perturbation Theory (APT) can be invoked instead and excitations predicted in terms of quasi-particles \cite{Polkovnikov1,*Polkovnikov,Viola}. However, totally connected lattice models like the Dicke Model (DM) and the Lipkin-Meshkov-Glick Model (LMGM) \cite{Dicke,*Lipkin}, have no spatial order parameter and hence lack a clear connection to existing theories such as KZM. This may explain the lack of general results to date for the adiabatic QPT regime, with the exception of bosonic excitation estimates in the DM using simplifying mean-field and rotating wave approximations \cite{Altland1,*Altland}, scaling of final excitations in the LMGM \cite{LipCaneva}, and dynamical characterizations of the QPT through a monochromatic modulation of the annealing parameter \cite{BastidasDicke,*BastidasLipkin}.

\begin{figure*}
  \includegraphics[width=0.97 \textwidth]{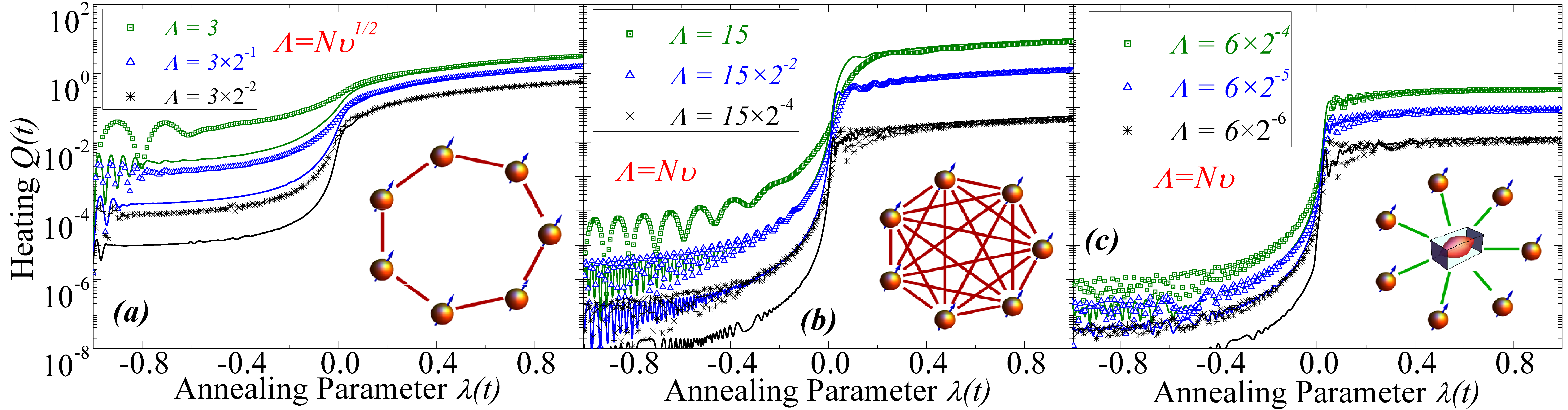}
  \caption{(color online) Time-evolution of heating $Q(t)$ for \textbf{(a)} TFIM, \textbf{(b)} LMGM, and \textbf{(c)} DM. Insets show how the qubits interact in these three models.
  In each panel, we present results for three different scaled annealing velocities $\Lambda$ (different colors) and two system sizes: TFIM, continuous lines ($N = 160$) and symbols ($N = 80$); LMGM, continuous lines ($N = 2^{11}$) and symbols ($N = 2^9$); DM, continuous lines ($N = 2^9$) and symbols ($N = 2^8$). Notice that for well inside the ordered phase ($\lambda>0$) for arbitrarily connected models, curves with the same $\Lambda$ but different size $N$ collapse.}\label{figQTotal}
\end{figure*}

  \begin{figure}
  \includegraphics[width=0.47 \textwidth]{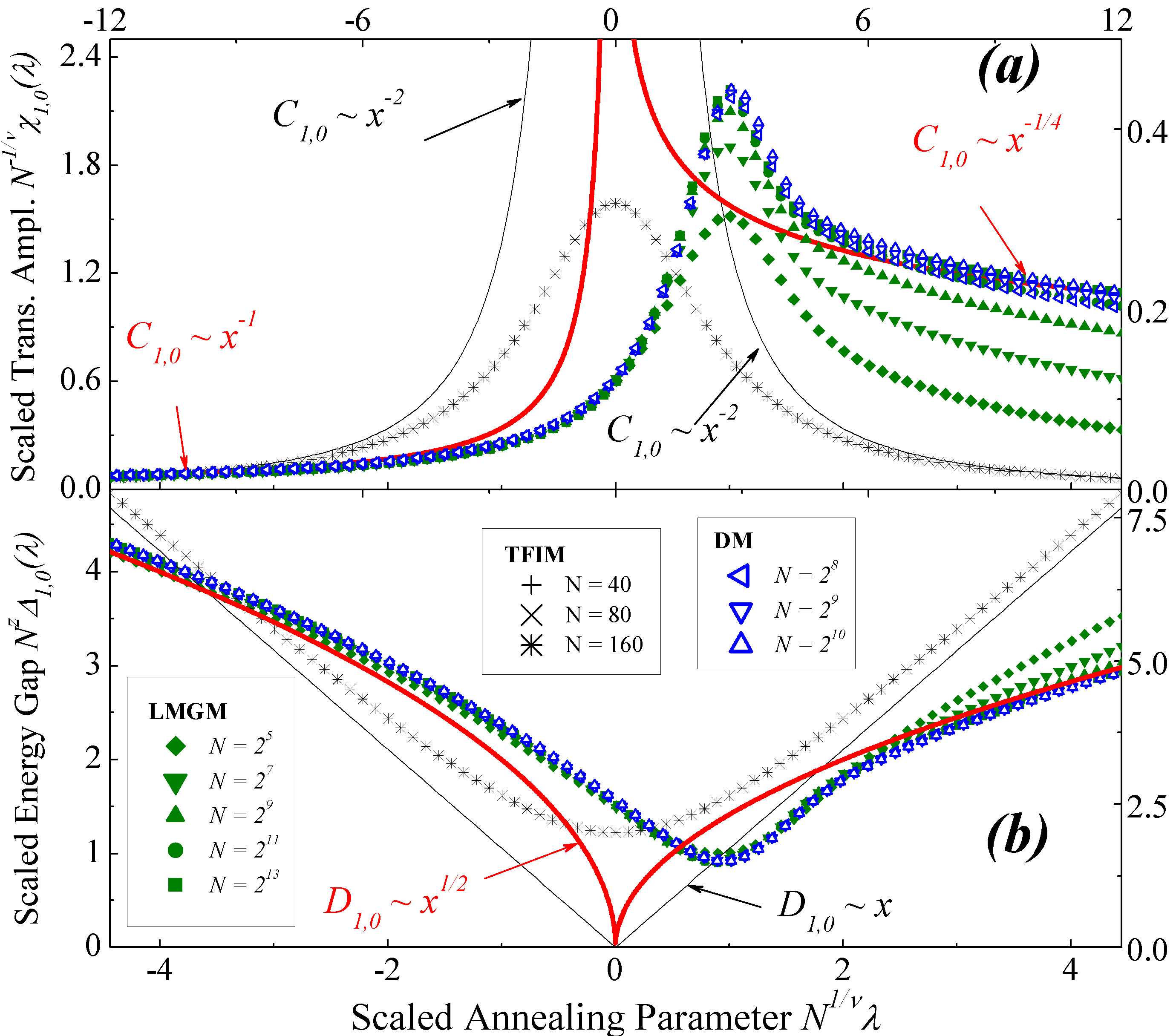}
  \caption{(color online) Universal behavior of finite-size critical functions \textbf{(a)} $C_{1,0}$ and \textbf{(b)} $D_{1,0}$ as defined in Eqs. \ref{eqchiFS} and \ref{eqDelFS}. Symbols show results for system size $N$ while continuous curves are power-law predictions in the TL. (Thick red line is for both DM and LMGM. Thin black line is for TFIM). Scales for the DM and LMGM are given at the left-bottom and right-top respectively, and show that the critical functions for both models have the same shape since they belong to the same universality class. Horizontal scale for the TFIM is at the top, while vertical scale is not present but goes up to $25$ in \textbf{(a)} and $0.25$ in \textbf{(b)}.}\label{figCritFun}
  \end{figure}

In a QPT, critical exponents are extracted from the power-law behavior in the thermodynamic limit (TL), of equilibrium quantities such as energy gaps and susceptibilities around the quantum critical point (QCP) \cite{Dziarmaga}. However when finite-size scaling is considered, continuous functions emerge at the phase boundary. Here we show that, contrary to common belief, these critical functions -- but not critical exponents -- provide a unified description of QPT dynamics, as encoded by nonadiabatic indicators like heating and ground state fidelity. Furthermore they encompass both finite-range (e.g. Transverse Field Ising Model (TFIM)) and fully-connected systems (e.g. DM and LMGM \cite{ReslenEPL}) and hence overcome the limitations of KZM and APT.
In addition to applications in adiabatic quantum computing \cite{AdQuCom1,*AdQuCom,Lidar1,*Lidar2}, QPTs have been experimentally realized using ultracold atomic systems \cite{Greiner,BlochRev}.
By revealing continuous-time details of the size-independent dynamical behavior of quantum many-body systems, our analysis goes beyond critical exponent analyses such as KZM and connects to studies of avalanche-like events across classical phase boundaries \cite{BeyondCrit}. KZM-like and an APT-like regime are both naturally incorporated in, and illuminated by, this new framework. We find that the traditional universality of critical exponents is insufficient to describe the analogous dynamical evolutions of different models, thereby casting doubt on an implicit assumption of quantum simulations \cite{QuantSimulEditr} in which particular models are taken to act as experimental surrogates of each other.

\begin{figure*}
  \includegraphics[width=0.97 \textwidth]{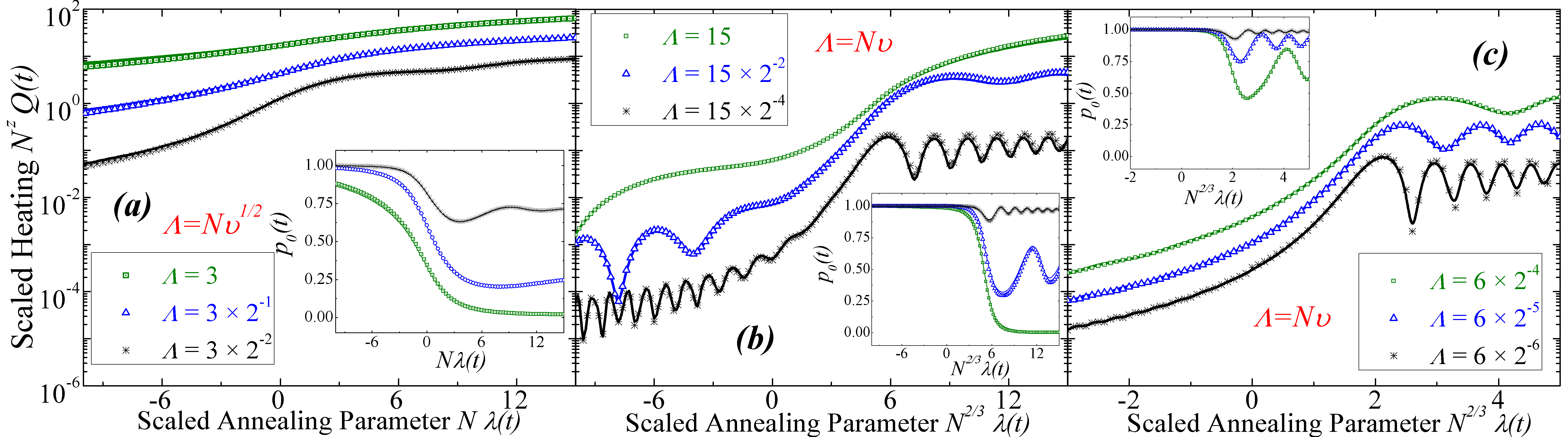}
  \caption{(color online) Magnification and rescaling of the results in Fig. \ref{figQTotal}  around QCP, as described by Eq.  \ref{eqtsimplefinita}. Collapse is no longer restricted to a nearly constant value well after the QCP. Instead, a continuous-in-time size-independent behavior is revealed, a prediction that lies outside the scope of critical exponent analysis. Insets show continuous scaling also present for the ground state fidelity $p_0 (t)$. Exponent $z=1$ for the TFIM, while $z=1/3$ for both LMGM and DM.}\label{figScaled}
\end{figure*}

We focus on three models for which experimental realizations exist or have been proposed  \cite{CiracPorras,*Baumann,*Simon}. Each features $N$ qubits with differing interaction connectivities (Fig. \ref{figQTotal} insets). The following generic, dimensionless, time-dependent Hamiltonian ($\hbar = 1$) describes the TFIM and LMGM:
\begin{equation} \hat{H} (t) =
-\sum_{i=1}^{N} \hat{\sigma}_{z}^{(i)}
 -\frac{\lambda(t)+1}{N^s}\sum_{\langle i,j\rangle}\hat{\sigma}_{x}^{(i)}\hat{\sigma}_{x}^{(j)},
\label{eqHdos}
\end{equation}
where $\{\hat{\sigma}\}$ are Pauli matrices and $\langle i,j\rangle$ denotes pairs of interacting qubits. We choose the interaction strength $\lambda (t)$ as the annealing parameter. In the TFIM, only pairs of nearest neighbors in a circular lattice interact, while in the LMGM all pair interactions are present. At $\lambda = 0$ in the TL, the system is at the QCP \cite{VidalLipkin,Sachdev}, where a minimum-value energy gap arises between the ground state and the first excited accessible state. $N^s$ in the denominator normalizes the interaction parameter according to size for the LMGM. $s=0$ in the TFIM, while $s=1$ in the LMGM. Like the LMGM, the DM has a totally connected lattice, but the qubit-qubit interaction is mediated by a boson mode: when the qubits and boson mode are in resonance, the Hamiltonian becomes:
\begin{equation} \hat{H} (t) = \sum_{i=1}^{N} \hat{\sigma}_{z}^{(i)}
+ \hat{a}^{\dag }\hat{a}+
\frac{\lambda (t) +1}{2 \sqrt{N}} \left( \hat{a} ^{\dag }+\hat{a}\right) \sum_{i=1}^{N} \hat{\sigma}_{x}^{(i)}
\label{eqHDM}
\end{equation}
where $\hat{a}^{\dag }$ ($\hat{a}$) is the mode's creation (annihilation)
operator. The QCP in the TL is also at $\lambda = 0$ \cite{BrandesPRL}.

To describe the crossing of the QPT, we show a simple case in which the annealing parameter evolves linearly as $\lambda(t) = \upsilon t$ since it leads to relatively simple formulae, though we stress that extension to any power-law time-dependence is straightforward \cite{Supplementary}. We start in the ground state $\left\vert \varphi_0(t_{\mathrm{i}}) \right\rangle$ with $\lambda(t_{\mathrm{i}})=-1$ at `negative' time $t_i=-\upsilon^{-1}$ (i.e. the zero of time is defined as the instance where the system passes through $\lambda(t_{\mathrm{i}})=0$). The systems in Eqs. \ref{eqHdos} and \ref{eqHDM} evolve with a time-dependent state $\left\vert \Psi (t) \right\rangle$ across the QCP, until positive time $t_{\mathrm{f}}$ where $\lambda(t_{\mathrm{f}})=1$. For slow enough quench, the system should end in a final ground state $\left\vert \varphi_0(\lambda(t_{\mathrm{f}})) \right\rangle$ representing perfect adiabatic evolution. However the QCP hinders the many-body system from achieving this result, since the minimal energy gap makes it easy for the system to jump out of the ground state. Because this gap gets smaller as the system size increases, ever slower quenches are necessary to keep the system in the ground state. Hence the fundamental effect of the crossing of QPTs is the loss of adiabatic evolution. We employ two indicators to probe this result: (1) Ground state fidelity
$p_0(t) = \left\vert \left\langle \varphi_0 (t) \vert \Psi (t) \right\rangle \right\vert ^2$ measuring the overlap between the actual dynamical state $\left\vert \Psi (t) \right\rangle$ and $\left\vert \varphi_0(t) \right\rangle$. It lies in the interval $0 \le p_0 \le 1$ and has its maximum value for perfect adiabatic evolution. (2) Heating $Q(t)=\left\langle \Psi (t) \right\vert \hat{H}(t)  \left\vert  \Psi (t) \right\rangle -E_0(t)$ which is always non-negative, and for adiabatic evolution is zero \cite{PolkovnikovNat}. ($E_0(t)$ is the instantaneous ground-state energy). Details of the calculation are shown in the Supplementary Material \cite{Supplementary}.

 Figure \ref{figQTotal} shows the heating $Q(t)$ for   $\lambda \in [-1,1]$. For $\lambda < 0$, the behavior at a given $\upsilon$ is independent of size, with virtually no loss of adiabaticity provided $\upsilon$ is small enough \cite{Polkovnikov}. The stronger heating behavior that emerges above the QCP results from excitations forming, following a scaled velocity $\Lambda$. The almost vertical step around $\lambda =0$ shows that the evolution is essentially adiabatic, except for the narrow interval around the QCP where the major excitations are formed. Since the important aspects of the quenching are defined around the QCP, we analyze the system's state in terms of instantaneous eigenstates:
\begin{equation}
\left\vert \Psi (t) \right\rangle = \sum _{n=0}  a_n (t) \mathrm{e}^{- i
\int _{t_0}^{t}  E_n (t ^{\prime}) \mathrm{d} t ^{\prime} } \left\vert \varphi_{\lambda (t)} ^{(n)}
\right\rangle,
\label{eq3}
\end{equation}
where $\hat{H}(t)
\left\vert \varphi_{\lambda(t)} ^{(n)} \right\rangle = E_n (\lambda(t)) \left\vert
\varphi_{\lambda(t)} ^{(n)} \right\rangle $ for every time $t$. The $a_n (t)$ evolution follows
\begin{equation}
\frac{\mathrm{d} a_n (\lambda)}{\mathrm{d} \lambda}    =     \sum _{m \neq n}
\mathrm{e}^{i \upsilon^{-1} \phi_{n,m}^{(N)}(\lambda)}\chi_{n,m} ^{(N)}(\lambda)  a_m (\lambda);
\label{eqt1}
\end{equation}
where $\phi_{n,m}^{(N)}(\lambda) \equiv \int _{0} ^{\lambda} \Delta_{n,m}^{(N)} (\lambda ^{\prime}) \mathrm{d} \lambda ^{\prime}$, which is the integral of the energy gap between eigenstates $n$ and $m$. The transition amplitudes $\chi_{n,m}^{(N)} \equiv - \left\langle  \varphi_{\lambda} ^{(n)}  \right\vert  \frac{\mathrm{d}}{\mathrm{d} \lambda} \left(   \left\vert  \varphi_{\lambda} ^{(m)} \right\rangle \right)$ can be written as $\chi_{n,m}^{(N)}(\lambda)= \frac{V_{n,m}^{(N)}(\lambda)}{\Delta_{n,m}^{(N)}(\lambda)}$ whenever eigenstates $n$ and $m$ are non-degenerate, and $V_{n,m}^{(N)}$ are the matrix elements of the interaction part of the Hamiltonian, mediated by $\lambda$. The superscript $(N)$ indicates that all the functions depend on the system size. Equation \ref{eqt1} is usually the central part of the Adiabatic Theorem \cite{Messiah}, which states that if $\upsilon \ll \left\vert \frac{\Delta_{n,m}}{\chi_{n,m}} \right\vert$, then $\{a_n\}$ remain constant. For sufficiently slow annealing, this is satisfied outside the QCP region, implying that only the eigenstates that reach a zero gap in the TL are relevant for the loss of adiabaticity. It is at this point that dynamical critical functions enter the picture, since $\chi_{n,m}$ and $\Delta_{n,m}$  obey a scaling relation when $\vert \lambda \vert \ll 1$ \cite{VidalLipkin,VidalDicke,Viola}:
\begin{eqnarray}
 \chi_{n,m}^{(N)}(\lambda) &=& N^{1/\nu} C_{n,m}(x),
 \label{eqchiFS}
 \\
 \Delta_{n,m}^{(N)}(\lambda) &=&  N^{-z} D_{n,m}(x);
 \label{eqDelFS}
\end{eqnarray}
where $x \equiv N^{1/\nu} \lambda $ with $\nu$ and $z$ determined through the power-law behavior in the TL \cite{BrandesPRL,Sachdev}. Figures \ref{figCritFun}(a),(b) present these critical functions between the ground and first-excited state, and show how they start matching the TL power-law behavior for sufficiently large $N$ and $x$. It follows that $\nu = z = 1$ for the TFIM, and $\nu = 3/2$ and $z = 1/3$ for DM and LMGM \cite{Botet,*BrandesPRE,Sachdev}.

Our main result is that the evolution in Eq. \ref{eqt1} can now be cast in size-independent form:
\begin{equation}
\frac{\mathrm{d} a_n (x)}{\mathrm{d} x} = \sum _{n \neq m} \mathrm{e}^{i \Lambda^{-1/\mu} \Phi_{n,m}(x)}   a_m(x) C_{n,m} (x),
\label{eqtsimplefinita}
\end{equation}
where the scaled velocity $\Lambda = N \upsilon^{\mu}$, the dynamical phase difference $\Phi_{n,m} (x) = \int_0^x D_{n,m} (x ^{\prime}) \mathrm{d} x ^{\prime}$, and $\mu = \frac{\nu}{1+z\nu}$. Equation \ref{eqtsimplefinita} predicts universal results in terms of excitation probabilities $p_n \equiv \vert a_n \vert^2$, and the ground-state fidelity. Since the energy spectrum has a regular behavior, the heating will also be universal since $Q \equiv \sum _n p_n \Delta_{n,0}$. This prediction is confirmed in Fig. \ref{figScaled} with both adiabatic quantifiers behaving in a size-independent manner across the critical region. We note that the collapse only occurs during and after the critical threshold, because it is around the QCP that the adiabatic indicators are significantly affected, and it is in this region that universal functions exist.  Once the QCP is passed ($\lambda>0$ stage), the evolution is again essentially adiabatic and the accumulated non-adiabatic effects of crossing the critical threshold remain dominant, clamping the subsequent collapsed evolution. Therefore, our results show that by generalizing from critical exponents to critical functions,  we expand the traditional description focused on scaling at a fixed final value of $\lambda$ as in Fig. \ref{figQTotal}, to a complete temporal collapse picture as shown in Fig. \ref{figScaled} around the QCP.

 This new picture includes well-known results predicted by KZM and APT as special cases, since both can be expressed as power-law dependencies at the end of the quenching process \cite{LipCaneva}. In the lower velocity APT regime where there is low probability of leaving the ground state, the following approximation holds for $n \ne 0$:
\begin{equation}
a_n (\lambda) \approx \int_{-1}^{\lambda} \mathrm{e}^{i \upsilon^{-1} \phi_{n,0}(\lambda ^{^{\prime}})}\chi_{n,0}(\lambda ^{^{\prime}}) \mathrm{d} \lambda ^{^{\prime}} \ .
\end{equation}
Since the integrand is only non-negligible around the QCP, it follows that $\vert a_n \vert \sim \upsilon$, for which $p_n \sim \upsilon^2$ and hence $Q_f \sim \upsilon^2$.
The higher velocity KZM regime is characterized by excitations being large enough to discard APT, replacing it by an adiabatic-impulse-adiabatic approximation in which the size of the threshold at which major excitations are created is defined by a definite time $t_K$, or equivalently a critical value of $x_K=\upsilon N^{1/\nu} t_K$.

In the traditional KZM, as is the case of the TFIM, a healing time has been directly related to the inverse energy gap \cite{Dziarmaga}, making $t_K \Delta(t_K) = t_k (\upsilon t_K)^{\nu z} \sim 1$ and then $x_K \sim \Lambda ^{1/\nu}$. With this estimate, an adiabatic indicator such as heating can be predicted through $Q_f \sim D(x_K) \sim \Lambda ^ z$.
For the TFIM ($z=1$), this KZM prediction has been confirmed \cite{DziZurPRL,*DziZur}. By stark contrast, totally connected models do not match this estimate: instead of an exponent $1/3$, a scaling $Q_f \sim \Lambda ^ {3/2} $ has been found \cite{LipCaneva}. However, Fig. \ref{figCritFun}(a) reveals that in the $\lambda > 0$ phase, there is an anomalous $x^{-1/4}$ dependence caused by a divergent $\chi \sim N^{1/2}$ transition amplitude \cite{BrandesPRE}. Furthermore in the $\lambda < 0$ phase, a $x^{-1}$ dependence is present. Such exponents are specific to infinite dimensional lattices like the LMGM and DM, and this difference is not taken into account in the KZM.

The failure of the KZM prediction for LMGM and DM, highlights the accuracy of a dynamical function approach as compared to power-law relations based on critical exponents. Dynamical critical functions provide a full time-resolved picture of dynamical scaling in the near-adiabatic regime, even around the critical threshold where excitations have not yet stabilized -- hence understanding their properties is crucial for the design and cross-checking of annealing schemes in quantum simulations. The fact that the curves for LMGM and DM in Figs. \ref{figCritFun}(a) and (b) have essentially the same shape, might erroneously be taken as sufficient justification for using one as a  quantum simulation of the other -- however, this is not true. No matter how a dynamical curve in Fig. \ref{figScaled}(b) is scaled, its shape will  never completely match any curve of Fig. \ref{figScaled}(c). Instead, a thorough examination of Eq. \ref{eqtsimplefinita} reveals that equivalence between both near-adiabatic evolutions can only be achieved if the functions $\{C_{n,m} (x)\}$ scale as  $C_{n,m}^{\mathrm{DM}} (\alpha x) = \alpha^{-1} C_{n,m}^{\mathrm{LMGM}} (x)$, which is a stringent condition that is undetectable through critical exponent analysis.
In short, although equilibrium equivalence between systems around the QCP can be accomplished just by having identical critical exponents, achieving {\em dynamical} equivalence requires further tuning of model parameters, thereby partitioning the traditionally static universality classes into {\em subsets} of dynamically equivalent systems.

\begin{acknowledgements}

The authors thank Ana Maria Rey for a critical reading of the manuscript. O.L.A, L.Q. and F.J.R. acknowledge financial support from Proyectos Semilla-Facultad de Ciencias at
Universidad de los Andes (2010-2012) and project {\it Quantum control of non-equilibrium hybrid
systems}, UniAndes-2013 . O.~L.~A. acknowledges financial support from Colciencias, Convocatoria
511.

\end{acknowledgements}

\bibliography{bibliography}

\begin{thebibliography}{40}%
\makeatletter
\providecommand \@ifxundefined [1]{%
 \@ifx{#1\undefined}
}%
\providecommand \@ifnum [1]{%
 \ifnum #1\expandafter \@firstoftwo
 \else \expandafter \@secondoftwo
 \fi
}%
\providecommand \@ifx [1]{%
 \ifx #1\expandafter \@firstoftwo
 \else \expandafter \@secondoftwo
 \fi
}%
\providecommand \natexlab [1]{#1}%
\providecommand \enquote  [1]{``#1''}%
\providecommand \bibnamefont  [1]{#1}%
\providecommand \bibfnamefont [1]{#1}%
\providecommand \citenamefont [1]{#1}%
\providecommand \href@noop [0]{\@secondoftwo}%
\providecommand \href [0]{\begingroup \@sanitize@url \@href}%
\providecommand \@href[1]{\@@startlink{#1}\@@href}%
\providecommand \@@href[1]{\endgroup#1\@@endlink}%
\providecommand \@sanitize@url [0]{\catcode `\\12\catcode `\$12\catcode
  `\&12\catcode `\#12\catcode `\^12\catcode `\_12\catcode `\%12\relax}%
\providecommand \@@startlink[1]{}%
\providecommand \@@endlink[0]{}%
\providecommand \url  [0]{\begingroup\@sanitize@url \@url }%
\providecommand \@url [1]{\endgroup\@href {#1}{\urlprefix }}%
\providecommand \urlprefix  [0]{URL }%
\providecommand \Eprint [0]{\href }%
\providecommand \doibase [0]{http://dx.doi.org/}%
\providecommand \selectlanguage [0]{\@gobble}%
\providecommand \bibinfo  [0]{\@secondoftwo}%
\providecommand \bibfield  [0]{\@secondoftwo}%
\providecommand \translation [1]{[#1]}%
\providecommand \BibitemOpen [0]{}%
\providecommand \bibitemStop [0]{}%
\providecommand \bibitemNoStop [0]{.\EOS\space}%
\providecommand \EOS [0]{\spacefactor3000\relax}%
\providecommand \BibitemShut  [1]{\csname bibitem#1\endcsname}%
\let\auto@bib@innerbib\@empty
\bibitem [{\citenamefont {Stanley}(1999)}]{Stanley}%
  \BibitemOpen
  \bibfield  {author} {\bibinfo {author} {\bibfnamefont {H.~E.}\ \bibnamefont
  {Stanley}},\ }\href {\doibase 10.1103/RevModPhys.71.S358} {\bibfield
  {journal} {\bibinfo  {journal} {Rev. Mod. Phys.}\ }\textbf {\bibinfo {volume}
  {71}},\ \bibinfo {pages} {S358} (\bibinfo {year} {1999})}\BibitemShut
  {NoStop}%
\bibitem [{\citenamefont {Papanikolaou}\ \emph {et~al.}(2011)\citenamefont
  {Papanikolaou}, \citenamefont {Bohn}, \citenamefont {Sommer}, \citenamefont
  {Durin}, \citenamefont {Zapperi},\ and\ \citenamefont {Sethna}}]{BeyondCrit}%
  \BibitemOpen
  \bibfield  {author} {\bibinfo {author} {\bibfnamefont {S.}~\bibnamefont
  {Papanikolaou}}, \bibinfo {author} {\bibfnamefont {F.}~\bibnamefont {Bohn}},
  \bibinfo {author} {\bibfnamefont {R.~L.}\ \bibnamefont {Sommer}}, \bibinfo
  {author} {\bibfnamefont {G.}~\bibnamefont {Durin}}, \bibinfo {author}
  {\bibfnamefont {S.}~\bibnamefont {Zapperi}}, \ and\ \bibinfo {author}
  {\bibfnamefont {J.~P.}\ \bibnamefont {Sethna}},\ }\href {\doibase
  10.1038/nphys1884} {\bibfield  {journal} {\bibinfo  {journal} {Nat. Phys.}\
  }\textbf {\bibinfo {volume} {7}},\ \bibinfo {pages} {316} (\bibinfo {year}
  {2011})}\BibitemShut {NoStop}%
\bibitem [{\citenamefont {Sachdev}(2011)}]{Sachdev}%
  \BibitemOpen
  \bibfield  {author} {\bibinfo {author} {\bibfnamefont {S.}~\bibnamefont
  {Sachdev}},\ }\href@noop {} {\emph {\bibinfo {title} {Quantum Phase
  Transitions}}}\ (\bibinfo  {publisher} {Cambridge University Press},\
  \bibinfo {year} {2011})\BibitemShut {NoStop}%
\bibitem [{\citenamefont {Farhi}\ \emph {et~al.}(2001)\citenamefont {Farhi},
  \citenamefont {Goldstone}, \citenamefont {Gutmann}, \citenamefont {Lapan},
  \citenamefont {Lundgren},\ and\ \citenamefont {Preda}}]{AdQuCom1}%
  \BibitemOpen
  \bibfield  {author} {\bibinfo {author} {\bibfnamefont {E.}~\bibnamefont
  {Farhi}}, \bibinfo {author} {\bibfnamefont {J.}~\bibnamefont {Goldstone}},
  \bibinfo {author} {\bibfnamefont {S.}~\bibnamefont {Gutmann}}, \bibinfo
  {author} {\bibfnamefont {J.}~\bibnamefont {Lapan}}, \bibinfo {author}
  {\bibfnamefont {A.}~\bibnamefont {Lundgren}}, \ and\ \bibinfo {author}
  {\bibfnamefont {D.}~\bibnamefont {Preda}},\ }\href {\doibase
  10.1126/science.1057726} {\bibfield  {journal} {\bibinfo  {journal}
  {Science}\ }\textbf {\bibinfo {volume} {292}},\ \bibinfo {pages} {472}
  (\bibinfo {year} {2001})}\BibitemShut {NoStop}%
\bibitem [{\citenamefont {Sch\"utzhold}\ and\ \citenamefont
  {Schaller}(2006)}]{AdQuCom}%
  \BibitemOpen
  \bibfield  {author} {\bibinfo {author} {\bibfnamefont {R.}~\bibnamefont
  {Sch\"utzhold}}\ and\ \bibinfo {author} {\bibfnamefont {G.}~\bibnamefont
  {Schaller}},\ }\href {\doibase 10.1103/PhysRevA.74.060304} {\bibfield
  {journal} {\bibinfo  {journal} {Phys. Rev. A}\ }\textbf {\bibinfo {volume}
  {74}},\ \bibinfo {pages} {060304} (\bibinfo {year} {2006})}\BibitemShut
  {NoStop}%
\bibitem [{\citenamefont {Boixo}\ \emph {et~al.}(2013)\citenamefont {Boixo},
  \citenamefont {R{\o}nnow}, \citenamefont {Isakov}, \citenamefont {Wang},
  \citenamefont {Wecker}, \citenamefont {Lidar}, \citenamefont {Martinis},\
  and\ \citenamefont {Troyer}}]{Lidar1}%
  \BibitemOpen
  \bibfield  {author} {\bibinfo {author} {\bibfnamefont {S.}~\bibnamefont
  {Boixo}}, \bibinfo {author} {\bibfnamefont {T.~F.}\ \bibnamefont
  {R{\o}nnow}}, \bibinfo {author} {\bibfnamefont {S.~V.}\ \bibnamefont
  {Isakov}}, \bibinfo {author} {\bibfnamefont {Z.}~\bibnamefont {Wang}},
  \bibinfo {author} {\bibfnamefont {D.}~\bibnamefont {Wecker}}, \bibinfo
  {author} {\bibfnamefont {D.~A.}\ \bibnamefont {Lidar}}, \bibinfo {author}
  {\bibfnamefont {J.~M.}\ \bibnamefont {Martinis}}, \ and\ \bibinfo {author}
  {\bibfnamefont {M.}~\bibnamefont {Troyer}},\ }\href
  {http://arxiv.org/pdf/1304.4595v2.pdf} {\  (\bibinfo {year} {2013})},\
  \Eprint {http://arxiv.org/abs/1304.4595} {arXiv:1304.4595 [quant-ph]}
  \BibitemShut {NoStop}%
\bibitem [{\citenamefont {Santra}\ \emph {et~al.}(2013)\citenamefont {Santra},
  \citenamefont {Quiroz}, \citenamefont {Steeg},\ and\ \citenamefont
  {Lidar}}]{Lidar2}%
  \BibitemOpen
  \bibfield  {author} {\bibinfo {author} {\bibfnamefont {S.}~\bibnamefont
  {Santra}}, \bibinfo {author} {\bibfnamefont {G.}~\bibnamefont {Quiroz}},
  \bibinfo {author} {\bibfnamefont {G.~V.}\ \bibnamefont {Steeg}}, \ and\
  \bibinfo {author} {\bibfnamefont {D.}~\bibnamefont {Lidar}},\ }\href
  {http://arxiv.org/pdf/1307.3931v1.pdf} {\  (\bibinfo {year} {2013})},\
  \Eprint {http://arxiv.org/abs/1307.3931} {arXiv:1307.3931 [quant-ph]}
  \BibitemShut {NoStop}%
\bibitem [{\citenamefont {Dziarmaga}(2010)}]{Dziarmaga}%
  \BibitemOpen
  \bibfield  {author} {\bibinfo {author} {\bibfnamefont {J.}~\bibnamefont
  {Dziarmaga}},\ }\href {\doibase 10.1080/00018732.2010.514702} {\bibfield
  {journal} {\bibinfo  {journal} {Advances in Physics}\ }\textbf {\bibinfo
  {volume} {59}},\ \bibinfo {pages} {1063} (\bibinfo {year}
  {2010})}\BibitemShut {NoStop}%
\bibitem [{\citenamefont {Zurek}\ \emph {et~al.}(2005)\citenamefont {Zurek},
  \citenamefont {Dorner},\ and\ \citenamefont {Zoller}}]{DziZurPRL}%
  \BibitemOpen
  \bibfield  {author} {\bibinfo {author} {\bibfnamefont {W.~H.}\ \bibnamefont
  {Zurek}}, \bibinfo {author} {\bibfnamefont {U.}~\bibnamefont {Dorner}}, \
  and\ \bibinfo {author} {\bibfnamefont {P.}~\bibnamefont {Zoller}},\ }\href
  {\doibase 10.1103/PhysRevLett.95.105701} {\bibfield  {journal} {\bibinfo
  {journal} {Phys. Rev. Lett.}\ }\textbf {\bibinfo {volume} {95}},\ \bibinfo
  {pages} {105701} (\bibinfo {year} {2005})}\BibitemShut {NoStop}%
\bibitem [{\citenamefont {Cincio}\ \emph {et~al.}(2007)\citenamefont {Cincio},
  \citenamefont {Dziarmaga}, \citenamefont {Rams},\ and\ \citenamefont
  {Zurek}}]{DziZur}%
  \BibitemOpen
  \bibfield  {author} {\bibinfo {author} {\bibfnamefont {L.}~\bibnamefont
  {Cincio}}, \bibinfo {author} {\bibfnamefont {J.}~\bibnamefont {Dziarmaga}},
  \bibinfo {author} {\bibfnamefont {M.~M.}\ \bibnamefont {Rams}}, \ and\
  \bibinfo {author} {\bibfnamefont {W.~H.}\ \bibnamefont {Zurek}},\ }\href
  {\doibase 10.1103/PhysRevA.75.052321} {\bibfield  {journal} {\bibinfo
  {journal} {Phys. Rev. A}\ }\textbf {\bibinfo {volume} {75}},\ \bibinfo
  {pages} {052321} (\bibinfo {year} {2007})}\BibitemShut {NoStop}%
\bibitem [{\citenamefont {Cucchietti}\ \emph {et~al.}(2007)\citenamefont
  {Cucchietti}, \citenamefont {Damski}, \citenamefont {Dziarmaga},\ and\
  \citenamefont {Zurek}}]{Cuchietti}%
  \BibitemOpen
  \bibfield  {author} {\bibinfo {author} {\bibfnamefont {F.~M.}\ \bibnamefont
  {Cucchietti}}, \bibinfo {author} {\bibfnamefont {B.}~\bibnamefont {Damski}},
  \bibinfo {author} {\bibfnamefont {J.}~\bibnamefont {Dziarmaga}}, \ and\
  \bibinfo {author} {\bibfnamefont {W.~H.}\ \bibnamefont {Zurek}},\ }\href
  {\doibase 10.1103/PhysRevA.75.023603} {\bibfield  {journal} {\bibinfo
  {journal} {Phys. Rev. A}\ }\textbf {\bibinfo {volume} {75}},\ \bibinfo
  {pages} {023603} (\bibinfo {year} {2007})}\BibitemShut {NoStop}%
\bibitem [{\citenamefont {Deng}\ \emph {et~al.}(2008)\citenamefont {Deng},
  \citenamefont {Ortiz},\ and\ \citenamefont {Viola}}]{Viola}%
  \BibitemOpen
  \bibfield  {author} {\bibinfo {author} {\bibfnamefont {S.}~\bibnamefont
  {Deng}}, \bibinfo {author} {\bibfnamefont {G.}~\bibnamefont {Ortiz}}, \ and\
  \bibinfo {author} {\bibfnamefont {L.}~\bibnamefont {Viola}},\ }\href
  {\doibase 10.1209/0295-5075/84/67008} {\bibfield  {journal} {\bibinfo
  {journal} {Europhys. Lett.}\ }\textbf {\bibinfo {volume} {84}},\ \bibinfo
  {pages} {67008} (\bibinfo {year} {2008})}\BibitemShut {NoStop}%
\bibitem [{\citenamefont {Polkovnikov}\ \emph {et~al.}(2011)\citenamefont
  {Polkovnikov}, \citenamefont {Sengupta}, \citenamefont {Silva},\ and\
  \citenamefont {Vengalattore}}]{TutDynAd}%
  \BibitemOpen
  \bibfield  {author} {\bibinfo {author} {\bibfnamefont {A.}~\bibnamefont
  {Polkovnikov}}, \bibinfo {author} {\bibfnamefont {K.}~\bibnamefont
  {Sengupta}}, \bibinfo {author} {\bibfnamefont {A.}~\bibnamefont {Silva}}, \
  and\ \bibinfo {author} {\bibfnamefont {M.}~\bibnamefont {Vengalattore}},\
  }\href {\doibase 10.1103/RevModPhys.83.863} {\bibfield  {journal} {\bibinfo
  {journal} {Rev. Mod. Phys.}\ }\textbf {\bibinfo {volume} {83}},\ \bibinfo
  {pages} {863} (\bibinfo {year} {2011})}\BibitemShut {NoStop}%
\bibitem [{\citenamefont {Kolodrubetz}\ \emph {et~al.}(2012)\citenamefont
  {Kolodrubetz}, \citenamefont {Clark},\ and\ \citenamefont {Huse}}]{Huse}%
  \BibitemOpen
  \bibfield  {author} {\bibinfo {author} {\bibfnamefont {M.}~\bibnamefont
  {Kolodrubetz}}, \bibinfo {author} {\bibfnamefont {B.~K.}\ \bibnamefont
  {Clark}}, \ and\ \bibinfo {author} {\bibfnamefont {D.~A.}\ \bibnamefont
  {Huse}},\ }\href {\doibase 10.1103/PhysRevLett.109.015701} {\bibfield
  {journal} {\bibinfo  {journal} {Phys. Rev. Lett.}\ }\textbf {\bibinfo
  {volume} {109}},\ \bibinfo {pages} {015701} (\bibinfo {year}
  {2012})}\BibitemShut {NoStop}%
\bibitem [{\citenamefont {Kibble}(1976)}]{Zurek2}%
  \BibitemOpen
  \bibfield  {author} {\bibinfo {author} {\bibfnamefont {T.}~\bibnamefont
  {Kibble}},\ }\href {\doibase 10.1088/0305-4470/9/8/029} {\bibfield  {journal}
  {\bibinfo  {journal} {J. of Phys. A: Math. Gen.}\ }\textbf {\bibinfo {volume}
  {9}},\ \bibinfo {pages} {1387} (\bibinfo {year} {1976})}\BibitemShut
  {NoStop}%
\bibitem [{\citenamefont {Zurek}(1985)}]{Zurek}%
  \BibitemOpen
  \bibfield  {author} {\bibinfo {author} {\bibfnamefont {W.~H.}\ \bibnamefont
  {Zurek}},\ }\href {\doibase 10.1038/317505a0} {\bibfield  {journal} {\bibinfo
   {journal} {Nature}\ }\textbf {\bibinfo {volume} {317}},\ \bibinfo {pages}
  {505} (\bibinfo {year} {1985})}\BibitemShut {NoStop}%
\bibitem [{\citenamefont {Polkovnikov}(2005)}]{Polkovnikov1}%
  \BibitemOpen
  \bibfield  {author} {\bibinfo {author} {\bibfnamefont {A.}~\bibnamefont
  {Polkovnikov}},\ }\href {\doibase 10.1103/PhysRevB.72.161201} {\bibfield
  {journal} {\bibinfo  {journal} {Phys. Rev. B}\ }\textbf {\bibinfo {volume}
  {72}},\ \bibinfo {pages} {161201} (\bibinfo {year} {2005})}\BibitemShut
  {NoStop}%
\bibitem [{\citenamefont {De~Grandi}\ \emph {et~al.}(2010)\citenamefont
  {De~Grandi}, \citenamefont {Gritsev},\ and\ \citenamefont
  {Polkovnikov}}]{Polkovnikov}%
  \BibitemOpen
  \bibfield  {author} {\bibinfo {author} {\bibfnamefont {C.}~\bibnamefont
  {De~Grandi}}, \bibinfo {author} {\bibfnamefont {V.}~\bibnamefont {Gritsev}},
  \ and\ \bibinfo {author} {\bibfnamefont {A.}~\bibnamefont {Polkovnikov}},\
  }\href {\doibase 10.1103/PhysRevB.81.012303} {\bibfield  {journal} {\bibinfo
  {journal} {Phys. Rev. B}\ }\textbf {\bibinfo {volume} {81}},\ \bibinfo
  {pages} {012303} (\bibinfo {year} {2010})}\BibitemShut {NoStop}%
\bibitem [{\citenamefont {Dicke}(1954)}]{Dicke}%
  \BibitemOpen
  \bibfield  {author} {\bibinfo {author} {\bibfnamefont {R.~H.}\ \bibnamefont
  {Dicke}},\ }\href {\doibase 10.1103/PhysRev.93.99} {\bibfield  {journal}
  {\bibinfo  {journal} {Phys. Rev.}\ }\textbf {\bibinfo {volume} {93}},\
  \bibinfo {pages} {99} (\bibinfo {year} {1954})}\BibitemShut {NoStop}%
\bibitem [{\citenamefont {Lipkin}\ \emph {et~al.}(1965)\citenamefont {Lipkin},
  \citenamefont {Meshkov},\ and\ \citenamefont {Glick}}]{Lipkin}%
  \BibitemOpen
  \bibfield  {author} {\bibinfo {author} {\bibfnamefont {H.}~\bibnamefont
  {Lipkin}}, \bibinfo {author} {\bibfnamefont {N.}~\bibnamefont {Meshkov}}, \
  and\ \bibinfo {author} {\bibfnamefont {A.}~\bibnamefont {Glick}},\ }\href
  {\doibase 10.1016/0029-5582(65)90862-X} {\bibfield  {journal} {\bibinfo
  {journal} {Nuclear Physics}\ }\textbf {\bibinfo {volume} {62}},\ \bibinfo
  {pages} {188 } (\bibinfo {year} {1965})}\BibitemShut {NoStop}%
\bibitem [{\citenamefont {Altland}\ and\ \citenamefont
  {Gurarie}(2008)}]{Altland1}%
  \BibitemOpen
  \bibfield  {author} {\bibinfo {author} {\bibfnamefont {A.}~\bibnamefont
  {Altland}}\ and\ \bibinfo {author} {\bibfnamefont {V.}~\bibnamefont
  {Gurarie}},\ }\href {\doibase 10.1103/PhysRevLett.100.063602} {\bibfield
  {journal} {\bibinfo  {journal} {Phys. Rev. Lett.}\ }\textbf {\bibinfo
  {volume} {100}},\ \bibinfo {pages} {063602} (\bibinfo {year}
  {2008})}\BibitemShut {NoStop}%
\bibitem [{\citenamefont {Altland}\ \emph {et~al.}(2009)\citenamefont
  {Altland}, \citenamefont {Gurarie}, \citenamefont {Kriecherbauer},\ and\
  \citenamefont {Polkovnikov}}]{Altland}%
  \BibitemOpen
  \bibfield  {author} {\bibinfo {author} {\bibfnamefont {A.}~\bibnamefont
  {Altland}}, \bibinfo {author} {\bibfnamefont {V.}~\bibnamefont {Gurarie}},
  \bibinfo {author} {\bibfnamefont {T.}~\bibnamefont {Kriecherbauer}}, \ and\
  \bibinfo {author} {\bibfnamefont {A.}~\bibnamefont {Polkovnikov}},\ }\href
  {\doibase 10.1103/PhysRevA.79.042703} {\bibfield  {journal} {\bibinfo
  {journal} {Phys. Rev. A}\ }\textbf {\bibinfo {volume} {79}},\ \bibinfo
  {pages} {042703} (\bibinfo {year} {2009})}\BibitemShut {NoStop}%
\bibitem [{\citenamefont {Caneva}\ \emph {et~al.}(2008)\citenamefont {Caneva},
  \citenamefont {Fazio},\ and\ \citenamefont {Santoro}}]{LipCaneva}%
  \BibitemOpen
  \bibfield  {author} {\bibinfo {author} {\bibfnamefont {T.}~\bibnamefont
  {Caneva}}, \bibinfo {author} {\bibfnamefont {R.}~\bibnamefont {Fazio}}, \
  and\ \bibinfo {author} {\bibfnamefont {G.~E.}\ \bibnamefont {Santoro}},\
  }\href {\doibase 10.1103/PhysRevB.78.104426} {\bibfield  {journal} {\bibinfo
  {journal} {Phys. Rev. B}\ }\textbf {\bibinfo {volume} {78}},\ \bibinfo
  {pages} {104426} (\bibinfo {year} {2008})}\BibitemShut {NoStop}%
\bibitem [{\citenamefont {Bastidas}\ \emph {et~al.}(2012)\citenamefont
  {Bastidas}, \citenamefont {Emary}, \citenamefont {Regler},\ and\
  \citenamefont {Brandes}}]{BastidasDicke}%
  \BibitemOpen
  \bibfield  {author} {\bibinfo {author} {\bibfnamefont {V.~M.}\ \bibnamefont
  {Bastidas}}, \bibinfo {author} {\bibfnamefont {C.}~\bibnamefont {Emary}},
  \bibinfo {author} {\bibfnamefont {B.}~\bibnamefont {Regler}}, \ and\ \bibinfo
  {author} {\bibfnamefont {T.}~\bibnamefont {Brandes}},\ }\href {\doibase
  10.1103/PhysRevLett.108.043003} {\bibfield  {journal} {\bibinfo  {journal}
  {Phys. Rev. Lett.}\ }\textbf {\bibinfo {volume} {108}},\ \bibinfo {pages}
  {043003} (\bibinfo {year} {2012})}\BibitemShut {NoStop}%
\bibitem [{\citenamefont {Engelhardt}\ \emph {et~al.}(2013)\citenamefont
  {Engelhardt}, \citenamefont {Bastidas}, \citenamefont {Emary},\ and\
  \citenamefont {Brandes}}]{BastidasLipkin}%
  \BibitemOpen
  \bibfield  {author} {\bibinfo {author} {\bibfnamefont {G.}~\bibnamefont
  {Engelhardt}}, \bibinfo {author} {\bibfnamefont {V.~M.}\ \bibnamefont
  {Bastidas}}, \bibinfo {author} {\bibfnamefont {C.}~\bibnamefont {Emary}}, \
  and\ \bibinfo {author} {\bibfnamefont {T.}~\bibnamefont {Brandes}},\ }\href
  {\doibase 10.1103/PhysRevE.87.052110} {\bibfield  {journal} {\bibinfo
  {journal} {Phys. Rev. E}\ }\textbf {\bibinfo {volume} {87}},\ \bibinfo
  {pages} {052110} (\bibinfo {year} {2013})}\BibitemShut {NoStop}%
\bibitem [{\citenamefont {Reslen}\ \emph {et~al.}(2005)\citenamefont {Reslen},
  \citenamefont {Quiroga},\ and\ \citenamefont {Johnson}}]{ReslenEPL}%
  \BibitemOpen
  \bibfield  {author} {\bibinfo {author} {\bibfnamefont {J.}~\bibnamefont
  {Reslen}}, \bibinfo {author} {\bibfnamefont {L.}~\bibnamefont {Quiroga}}, \
  and\ \bibinfo {author} {\bibfnamefont {N.~F.}\ \bibnamefont {Johnson}},\
  }\href {\doibase 10.1209/epl/i2004-10313-4} {\bibfield  {journal} {\bibinfo
  {journal} {Europhys. Lett.}\ }\textbf {\bibinfo {volume} {69}},\ \bibinfo
  {pages} {8} (\bibinfo {year} {2005})}\BibitemShut {NoStop}%
\bibitem [{\citenamefont {Greiner}\ \emph {et~al.}(2002)\citenamefont
  {Greiner}, \citenamefont {Mandel}, \citenamefont {Esslinger}, \citenamefont
  {Hansch},\ and\ \citenamefont {Bloch}}]{Greiner}%
  \BibitemOpen
  \bibfield  {author} {\bibinfo {author} {\bibfnamefont {M.}~\bibnamefont
  {Greiner}}, \bibinfo {author} {\bibfnamefont {O.}~\bibnamefont {Mandel}},
  \bibinfo {author} {\bibfnamefont {T.}~\bibnamefont {Esslinger}}, \bibinfo
  {author} {\bibfnamefont {T.~W.}\ \bibnamefont {Hansch}}, \ and\ \bibinfo
  {author} {\bibfnamefont {I.}~\bibnamefont {Bloch}},\ }\href {\doibase
  10.1038/415039a} {\bibfield  {journal} {\bibinfo  {journal} {Nature}\
  }\textbf {\bibinfo {volume} {415}},\ \bibinfo {pages} {39} (\bibinfo {year}
  {2002})}\BibitemShut {NoStop}%
\bibitem [{\citenamefont {Bloch}\ \emph {et~al.}(2008)\citenamefont {Bloch},
  \citenamefont {Dalibard},\ and\ \citenamefont {Zwerger}}]{BlochRev}%
  \BibitemOpen
  \bibfield  {author} {\bibinfo {author} {\bibfnamefont {I.}~\bibnamefont
  {Bloch}}, \bibinfo {author} {\bibfnamefont {J.}~\bibnamefont {Dalibard}}, \
  and\ \bibinfo {author} {\bibfnamefont {W.}~\bibnamefont {Zwerger}},\ }\href
  {\doibase 10.1103/RevModPhys.80.885} {\bibfield  {journal} {\bibinfo
  {journal} {Rev. Mod. Phys.}\ }\textbf {\bibinfo {volume} {80}},\ \bibinfo
  {pages} {885} (\bibinfo {year} {2008})}\BibitemShut {NoStop}%
\bibitem [{\citenamefont {Trabesinger}(2012)}]{QuantSimulEditr}%
  \BibitemOpen
  \bibfield  {author} {\bibinfo {author} {\bibfnamefont {A.}~\bibnamefont
  {Trabesinger}},\ }\href {\doibase 10.1038/nphys2258} {\bibfield  {journal}
  {\bibinfo  {journal} {Nat. Phys.}\ }\textbf {\bibinfo {volume} {8}},\
  \bibinfo {pages} {263} (\bibinfo {year} {2012})}\BibitemShut {NoStop}%
\bibitem [{\citenamefont {Porras}\ and\ \citenamefont
  {Cirac}(2004)}]{CiracPorras}%
  \BibitemOpen
  \bibfield  {author} {\bibinfo {author} {\bibfnamefont {D.}~\bibnamefont
  {Porras}}\ and\ \bibinfo {author} {\bibfnamefont {J.~I.}\ \bibnamefont
  {Cirac}},\ }\href {\doibase 10.1103/PhysRevLett.92.207901} {\bibfield
  {journal} {\bibinfo  {journal} {Phys. Rev. Lett.}\ }\textbf {\bibinfo
  {volume} {92}},\ \bibinfo {pages} {207901} (\bibinfo {year}
  {2004})}\BibitemShut {NoStop}%
\bibitem [{\citenamefont {Baumann}\ \emph {et~al.}(2010)\citenamefont
  {Baumann}, \citenamefont {Guerlin}, \citenamefont {Brennecke},\ and\
  \citenamefont {Esslinger}}]{Baumann}%
  \BibitemOpen
  \bibfield  {author} {\bibinfo {author} {\bibfnamefont {K.}~\bibnamefont
  {Baumann}}, \bibinfo {author} {\bibfnamefont {C.}~\bibnamefont {Guerlin}},
  \bibinfo {author} {\bibfnamefont {F.}~\bibnamefont {Brennecke}}, \ and\
  \bibinfo {author} {\bibfnamefont {T.}~\bibnamefont {Esslinger}},\ }\href
  {\doibase 10.1038/nature09009} {\bibfield  {journal} {\bibinfo  {journal}
  {Nature}\ }\textbf {\bibinfo {volume} {464}},\ \bibinfo {pages} {1301}
  (\bibinfo {year} {2010})}\BibitemShut {NoStop}%
\bibitem [{\citenamefont {Simon}\ \emph {et~al.}(2011)\citenamefont {Simon},
  \citenamefont {Bakr}, \citenamefont {Ma}, \citenamefont {Tai}, \citenamefont
  {Preiss},\ and\ \citenamefont {Greiner}}]{Simon}%
  \BibitemOpen
  \bibfield  {author} {\bibinfo {author} {\bibfnamefont {J.}~\bibnamefont
  {Simon}}, \bibinfo {author} {\bibfnamefont {W.~S.}\ \bibnamefont {Bakr}},
  \bibinfo {author} {\bibfnamefont {R.}~\bibnamefont {Ma}}, \bibinfo {author}
  {\bibfnamefont {M.~E.}\ \bibnamefont {Tai}}, \bibinfo {author} {\bibfnamefont
  {P.~M.}\ \bibnamefont {Preiss}}, \ and\ \bibinfo {author} {\bibfnamefont
  {M.}~\bibnamefont {Greiner}},\ }\href {\doibase 10.1038/nature09994}
  {\bibfield  {journal} {\bibinfo  {journal} {Nature}\ }\textbf {\bibinfo
  {volume} {472}},\ \bibinfo {pages} {307} (\bibinfo {year}
  {2011})}\BibitemShut {NoStop}%
\bibitem [{\citenamefont {Dusuel}\ and\ \citenamefont
  {Vidal}(2004)}]{VidalLipkin}%
  \BibitemOpen
  \bibfield  {author} {\bibinfo {author} {\bibfnamefont {S.}~\bibnamefont
  {Dusuel}}\ and\ \bibinfo {author} {\bibfnamefont {J.}~\bibnamefont {Vidal}},\
  }\href {\doibase 10.1103/PhysRevLett.93.237204} {\bibfield  {journal}
  {\bibinfo  {journal} {Phys. Rev. Lett.}\ }\textbf {\bibinfo {volume} {93}},\
  \bibinfo {pages} {237204} (\bibinfo {year} {2004})}\BibitemShut {NoStop}%
\bibitem [{\citenamefont {Emary}\ and\ \citenamefont
  {Brandes}(2003{\natexlab{a}})}]{BrandesPRL}%
  \BibitemOpen
  \bibfield  {author} {\bibinfo {author} {\bibfnamefont {C.}~\bibnamefont
  {Emary}}\ and\ \bibinfo {author} {\bibfnamefont {T.}~\bibnamefont
  {Brandes}},\ }\href {\doibase 10.1103/PhysRevLett.90.044101} {\bibfield
  {journal} {\bibinfo  {journal} {Phys. Rev. Lett.}\ }\textbf {\bibinfo
  {volume} {90}},\ \bibinfo {pages} {044101} (\bibinfo {year}
  {2003}{\natexlab{a}})}\BibitemShut {NoStop}%
\bibitem [{Sup()}]{Supplementary}%
  \BibitemOpen
  \href@noop {} {}\bibinfo {note} {See supplementary material at [URL will be
  inserted by AIP].}\BibitemShut {Stop}%
\bibitem [{\citenamefont {Polkovnikov}\ and\ \citenamefont
  {Gritsev}(2008)}]{PolkovnikovNat}%
  \BibitemOpen
  \bibfield  {author} {\bibinfo {author} {\bibfnamefont {A.}~\bibnamefont
  {Polkovnikov}}\ and\ \bibinfo {author} {\bibfnamefont {V.}~\bibnamefont
  {Gritsev}},\ }\href {\doibase 10.1038/nphys963} {\bibfield  {journal}
  {\bibinfo  {journal} {Nat. Phys.}\ }\textbf {\bibinfo {volume} {4}},\
  \bibinfo {pages} {477} (\bibinfo {year} {2008})}\BibitemShut {NoStop}%
\bibitem [{\citenamefont {Messiah}(1981)}]{Messiah}%
  \BibitemOpen
  \bibfield  {author} {\bibinfo {author} {\bibfnamefont {A.}~\bibnamefont
  {Messiah}},\ }\href@noop {} {\emph {\bibinfo {title} {Quantum Mechanics}}},\
  Vol.~\bibinfo {volume} {2}\ (\bibinfo  {publisher} {North Holland},\ \bibinfo
  {year} {1981})\BibitemShut {NoStop}%
\bibitem [{\citenamefont {Vidal}\ and\ \citenamefont
  {Dusuel}(2006)}]{VidalDicke}%
  \BibitemOpen
  \bibfield  {author} {\bibinfo {author} {\bibfnamefont {J.}~\bibnamefont
  {Vidal}}\ and\ \bibinfo {author} {\bibfnamefont {S.}~\bibnamefont {Dusuel}},\
  }\href {\doibase 10.1209/epl/i2006-10041-9} {\bibfield  {journal} {\bibinfo
  {journal} {Europhys. Lett.}\ }\textbf {\bibinfo {volume} {74}},\ \bibinfo
  {pages} {817} (\bibinfo {year} {2006})}\BibitemShut {NoStop}%
\bibitem [{\citenamefont {Botet}\ \emph {et~al.}(1982)\citenamefont {Botet},
  \citenamefont {Jullien},\ and\ \citenamefont {Pfeuty}}]{Botet}%
  \BibitemOpen
  \bibfield  {author} {\bibinfo {author} {\bibfnamefont {R.}~\bibnamefont
  {Botet}}, \bibinfo {author} {\bibfnamefont {R.}~\bibnamefont {Jullien}}, \
  and\ \bibinfo {author} {\bibfnamefont {P.}~\bibnamefont {Pfeuty}},\ }\href
  {\doibase 10.1103/PhysRevLett.49.478} {\bibfield  {journal} {\bibinfo
  {journal} {Phys. Rev. Lett.}\ }\textbf {\bibinfo {volume} {49}},\ \bibinfo
  {pages} {478} (\bibinfo {year} {1982})}\BibitemShut {NoStop}%
\bibitem [{\citenamefont {Emary}\ and\ \citenamefont
  {Brandes}(2003{\natexlab{b}})}]{BrandesPRE}%
  \BibitemOpen
  \bibfield  {author} {\bibinfo {author} {\bibfnamefont {C.}~\bibnamefont
  {Emary}}\ and\ \bibinfo {author} {\bibfnamefont {T.}~\bibnamefont
  {Brandes}},\ }\href {\doibase 10.1103/PhysRevE.67.066203} {\bibfield
  {journal} {\bibinfo  {journal} {Phys. Rev. E}\ }\textbf {\bibinfo {volume}
  {67}},\ \bibinfo {pages} {066203} (\bibinfo {year}
  {2003}{\natexlab{b}})}\BibitemShut {NoStop}%
\end{thebibliography}%


\begin{thebibliography}{6}%
\makeatletter
\providecommand \@ifxundefined [1]{%
 \@ifx{#1\undefined}
}%
\providecommand \@ifnum [1]{%
 \ifnum #1\expandafter \@firstoftwo
 \else \expandafter \@secondoftwo
 \fi
}%
\providecommand \@ifx [1]{%
 \ifx #1\expandafter \@firstoftwo
 \else \expandafter \@secondoftwo
 \fi
}%
\providecommand \natexlab [1]{#1}%
\providecommand \enquote  [1]{``#1''}%
\providecommand \bibnamefont  [1]{#1}%
\providecommand \bibfnamefont [1]{#1}%
\providecommand \citenamefont [1]{#1}%
\providecommand \href@noop [0]{\@secondoftwo}%
\providecommand \href [0]{\begingroup \@sanitize@url \@href}%
\providecommand \@href[1]{\@@startlink{#1}\@@href}%
\providecommand \@@href[1]{\endgroup#1\@@endlink}%
\providecommand \@sanitize@url [0]{\catcode `\\12\catcode `\$12\catcode
  `\&12\catcode `\#12\catcode `\^12\catcode `\_12\catcode `\%12\relax}%
\providecommand \@@startlink[1]{}%
\providecommand \@@endlink[0]{}%
\providecommand \url  [0]{\begingroup\@sanitize@url \@url }%
\providecommand \@url [1]{\endgroup\@href {#1}{\urlprefix }}%
\providecommand \urlprefix  [0]{URL }%
\providecommand \Eprint [0]{\href }%
\providecommand \doibase [0]{http://dx.doi.org/}%
\providecommand \selectlanguage [0]{\@gobble}%
\providecommand \bibinfo  [0]{\@secondoftwo}%
\providecommand \bibfield  [0]{\@secondoftwo}%
\providecommand \translation [1]{[#1]}%
\providecommand \BibitemOpen [0]{}%
\providecommand \bibitemStop [0]{}%
\providecommand \bibitemNoStop [0]{.\EOS\space}%
\providecommand \EOS [0]{\spacefactor3000\relax}%
\providecommand \BibitemShut  [1]{\csname bibitem#1\endcsname}%
\let\auto@bib@innerbib\@empty
\bibitem [{\citenamefont {Zurek}\ \emph {et~al.}(2005)\citenamefont {Zurek},
  \citenamefont {Dorner},\ and\ \citenamefont {Zoller}}]{DziZurPRL}%
  \BibitemOpen
  \bibfield  {author} {\bibinfo {author} {\bibfnamefont {W.~H.}\ \bibnamefont
  {Zurek}}, \bibinfo {author} {\bibfnamefont {U.}~\bibnamefont {Dorner}}, \
  and\ \bibinfo {author} {\bibfnamefont {P.}~\bibnamefont {Zoller}},\ }\href
  {\doibase 10.1103/PhysRevLett.95.105701} {\bibfield  {journal} {\bibinfo
  {journal} {Phys. Rev. Lett.}\ }\textbf {\bibinfo {volume} {95}},\ \bibinfo
  {pages} {105701} (\bibinfo {year} {2005})}\BibitemShut {NoStop}%
\bibitem [{\citenamefont {Kolodrubetz}\ \emph {et~al.}(2012)\citenamefont
  {Kolodrubetz}, \citenamefont {Clark},\ and\ \citenamefont {Huse}}]{Huse}%
  \BibitemOpen
  \bibfield  {author} {\bibinfo {author} {\bibfnamefont {M.}~\bibnamefont
  {Kolodrubetz}}, \bibinfo {author} {\bibfnamefont {B.~K.}\ \bibnamefont
  {Clark}}, \ and\ \bibinfo {author} {\bibfnamefont {D.~A.}\ \bibnamefont
  {Huse}},\ }\href {\doibase 10.1103/PhysRevLett.109.015701} {\bibfield
  {journal} {\bibinfo  {journal} {Phys. Rev. Lett.}\ }\textbf {\bibinfo
  {volume} {109}},\ \bibinfo {pages} {015701} (\bibinfo {year}
  {2012})}\BibitemShut {NoStop}%
\bibitem [{\citenamefont {Zener}(1932)}]{LandauZener}%
  \BibitemOpen
  \bibfield  {author} {\bibinfo {author} {\bibfnamefont {C.}~\bibnamefont
  {Zener}},\ }\href {\doibase 10.1098/rspa.1932.0165} {\bibfield  {journal}
  {\bibinfo  {journal} {Proc. R. Soc. Lond. A}\ }\textbf {\bibinfo {volume}
  {137}},\ \bibinfo {pages} {696} (\bibinfo {year} {1932})}\BibitemShut
  {NoStop}%
\bibitem [{\citenamefont {Dusuel}\ and\ \citenamefont
  {Vidal}(2004)}]{VidalLipkin}%
  \BibitemOpen
  \bibfield  {author} {\bibinfo {author} {\bibfnamefont {S.}~\bibnamefont
  {Dusuel}}\ and\ \bibinfo {author} {\bibfnamefont {J.}~\bibnamefont {Vidal}},\
  }\href {\doibase 10.1103/PhysRevLett.93.237204} {\bibfield  {journal}
  {\bibinfo  {journal} {Phys. Rev. Lett.}\ }\textbf {\bibinfo {volume} {93}},\
  \bibinfo {pages} {237204} (\bibinfo {year} {2004})}\BibitemShut {NoStop}%
\bibitem [{\citenamefont {Emary}\ and\ \citenamefont
  {Brandes}(2003)}]{BrandesPRE}%
  \BibitemOpen
  \bibfield  {author} {\bibinfo {author} {\bibfnamefont {C.}~\bibnamefont
  {Emary}}\ and\ \bibinfo {author} {\bibfnamefont {T.}~\bibnamefont
  {Brandes}},\ }\href {\doibase 10.1103/PhysRevE.67.066203} {\bibfield
  {journal} {\bibinfo  {journal} {Phys. Rev. E}\ }\textbf {\bibinfo {volume}
  {67}},\ \bibinfo {pages} {066203} (\bibinfo {year} {2003})}\BibitemShut
  {NoStop}%
\bibitem [{\citenamefont {Chen}\ \emph {et~al.}(2008)\citenamefont {Chen},
  \citenamefont {Zhang}, \citenamefont {Liu},\ and\ \citenamefont
  {Wang}}]{ChenDicke}%
  \BibitemOpen
  \bibfield  {author} {\bibinfo {author} {\bibfnamefont {Q.-H.}\ \bibnamefont
  {Chen}}, \bibinfo {author} {\bibfnamefont {Y.-Y.}\ \bibnamefont {Zhang}},
  \bibinfo {author} {\bibfnamefont {T.}~\bibnamefont {Liu}}, \ and\ \bibinfo
  {author} {\bibfnamefont {K.-L.}\ \bibnamefont {Wang}},\ }\href {\doibase
  10.1103/PhysRevA.78.051801} {\bibfield  {journal} {\bibinfo  {journal} {Phys.
  Rev. A}\ }\textbf {\bibinfo {volume} {78}} (\bibinfo {year} {2008}),\
  10.1103/PhysRevA.78.051801}\BibitemShut {NoStop}%
\end{thebibliography}%

\end{document}


\title{Supplementary material to ``New dynamical scaling universality for quantum networks across adiabatic quantum phase transitions''}

\author{O.~L. Acevedo}
\author{L. Quiroga}
\author{F.~J. Rodr\'{i}guez}
\affiliation{Departamento de F\'{i}sica, Universidad de los Andes, A.A. 4976, Bogot\'{a}, Colombia}
\author{N.~F. Johnson}
\affiliation{Department of Physics, University of Miami, Coral Gables, Miami, FL 33124, USA}

\begin{abstract}
In this Supplementary Material (SM), we provide details of the computational strategies employed in the main paper, in order to obtain exact numerical results for adiabatically crossing quantum phase transitions (QPT) using finite-size versions of the Dicke Model (DM), Lipkin-Meshkov-Glick Model (LMGM) and Transverse-Field Ising Model (TFIM). These results can be extended in a relatively straightforward way to the case of a non-linear time-dependence of the annealing parameter.
\end{abstract}

\maketitle

\section{Solution of the TFIM problem}

The TFIM has the advantage of being integrable \cite{DziZurPRL}. After a Jordan-Wigner transformation, and a Fourier transform \cite{Huse}, its Hamiltonian can be decomposed as a sum of two-dimensional block Hamiltonians $\hat{H}(t) = \sum_k \hat{H}_k (t)$, where,
\begin{equation}
\hat{H}_k (t)= \hat{\sigma}_z^{(k)} + \left[ 1-\lambda(t)\right](\cos k \, \hat{\sigma}_z^{(k)}+\sin k \,  \hat{\sigma}_x^{(k)}),
\label{eqIsing1}
\end{equation}
 and $k=\frac{\pi}{N}(2n+1)$, with $n = 0,1,\ldots,N/2-1$. Mode $k$ represents a pair of fermionic excitations $\{k,-k\}$. These excitations are generated in even numbers because of parity conservation, where the operator is $\hat{\Pi} = \bigotimes _{i=1}^{N} \hat{\sigma}_z^{(i)}$. The decomposition in Eq. \ref{eqIsing1} means that instead of solving a $2^N$-dimensional Schrodinger equation, only $N/2$ two-dimensional equations of the form $\frac{\mathrm{d}}{\mathrm{d}t}\left\vert \psi _k (t)\right\rangle = -i \hat{H}_k (t) \left\vert \psi _k (t)\right\rangle$ need to be solved numerically, which are essentially Landau-Zener problems \cite{LandauZener}. The total solution will be just the direct product $\left\vert \Psi (t) \right\rangle= \bigotimes_k \left\vert \psi_k (t) \right\rangle$. Furthermore, total heating can be expressed as the sum of the heating for each independent solution $Q(t) =\sum_k Q ^{(k)} (t)$, while total fidelity is expressed as a product $p_0(t) = \prod _k p_0^{(k)} (t)$. As Eq. \ref{eqIsing1} allows exact diagonalization, exact forms for dynamical critical functions are possible and so the first-excited energy gap and transition amplitude are:
  \begin{eqnarray}
 \Delta_{1,0}^{(N)}(t) & = & 2\sqrt{\left[\lambda(t)+\cos (\pi/N)-1\right]^2+\sin^2 (\pi/N)}, \\
 \chi_{1,0}^{(N)}(t) & = & \frac{\sin (\pi/N) /2}{\left[\lambda(t)+\cos (\pi/N)-1\right]^2+\sin^2 (\pi/N)}.
 \end{eqnarray}

\section{Solution of the LMGM problem}

The LMGM is a non-integrable model, so it does not have an exact diagonalization, in contrast to the TFIM in the previous section. However, its conserved quantities can be exploited in order to greatly simplify its numerical simulation. As in the TFIM, parity is constant, and in addition total angular momentum $\boldsymbol{\hat{J}}^2$ is conserved \cite{VidalLipkin} in the sense of collective operators $\hat{J}_{i}=\frac{1}{2}\sum_{j=1}^{n_q}\hat{\sigma} _{i}^{\left( j\right) }$. This last symmetry permits us to cast the Hamiltonian in the form:
\begin{equation}
\hat{H}_(t) =-2\hat{J}_{z}- 4\frac{\lambda(t) +1}{N} \hat{J}_{x}^2.
\label{eqHLipkin}
\end{equation}
The even parity ($\Pi =1$), maximum angular momentum subspace ($J= N/2$) will be the one containing the entire adiabatic evolution of the system. In this sense, despite the total Hilbert space of the LMGM growing exponentially as $2^N$, the effective Hamiltonian is of the order $N/2$. In the basis of the even-parity eigenstates of $\hat{J}_z$, the Hamiltonian in Eq. \ref{eqHLipkin} is  bidiagonal which renders feasible the handling of systems sizes up to $N = 2^{13}$. Both dynamical evolution and instantaneous eigenstates were obtained with this basis.

\section{Solution of the DM problem}

The DM solution is significatively harder to obtain. It has, as the LMGM, total angular momentum as a conserved quantity, and again the dynamics will lie in the $J = N/2$ subspace which allows its Hamiltonian to be be expressed as:
\begin{equation}
\hat{H}(t)=\hat{J}_{z}+\hat{a}^{\dag }\hat{a}+
\frac{\lambda(t) +1}{\sqrt{N}} \hat{J}_{x}\left( \hat{a} ^{\dag }+\hat{a}\right).
\label{eqHDick}
\end{equation}
 The DM has also parity $\hat{\Pi} = \mathrm{e}^{i \pi \left(\hat{a}^{\dag}\hat{a}+\hat{J}_z-N/2\right)}$ as a conserved quantity \cite{BrandesPRE}. However, the DM poses more computational complications: besides being non-integrable, this model lies on an infinite-dimensional Hilbert space, and its conserved quantities do not decompose its Hamiltonian into a sum of finite-dimensional blocks. The traditional computational solution to this difficulty is to work with \emph{big enough} truncated versions of the Hilbert space. For adiabatic evolution situations like the ones we are interested in, this approach can be successful since the state-vectors of interest virtually \emph{lie} on finite subspaces, i.e., their projections onto these subspaces are almost identical to the state-vectors themselves. The success of this solution can be easily tested by extending the truncation limit and then noticing that the results do not change (convergence test). The naivest application of this solution to the DM would be to work with vectors of the form $\left\vert m \right\rangle_z \otimes \left\vert n \right\rangle$, where the first one is an eigenvector of $\hat{J}_z$ and the last one is a bosonic Fock state. The problem with this first approach is that it is not cost-efficient, especially when analyzing large systems during the symmetry-breaking phase ($\lambda > 0$). This is because during that phase, the ground state has an expectation value of the number operator $\hat{a}^{\dag}\hat{a}$ that grows with the annealing parameter $\lambda (t)$ and the number of qubits $N$, therefore requiring more Fock states to adequately represent the dynamics.
\\
\\
To confront that problem, Chen et al. have proposed a very useful basis that adapts itself to the behavior of the system as $\lambda$ changes \cite{ChenDicke}. This basis has shown spectacular performances for calculating finite-size static properties of the ground-state of the system, even when $N$ is very large. We have adapted this basis in order to take into account the conservation of parity. We find an enormous improvement in computational performance with respect to the basis from Fock states. If we define $G(t)=N^{-1/2}\left[1+\lambda(t)\right]$, the adapted vectors that form a basis for the Hilbert space have the form:
\begin{equation}
\left\vert \Phi _{m,k}\left( G\right) \right\rangle
=\frac{1}{\sqrt{2}}\left( \left\vert m\right\rangle_x \left\vert \varphi
_{m,k}\left(G\right)\right\rangle + \left( -1\right) ^{k}\left\vert -m\right\rangle_x
\left\vert \varphi _{-m,k}\left(G\right)\right\rangle \right). \label{eqDickbasis}
\end{equation}
 Here the states $\left\vert m\right\rangle_x$ are eigenvectors of the $\hat{J}_x$ operator with eigenvalue $m$. The states $\left\vert \varphi_{m,k}\left(G\right)\right\rangle$ would be eigenvectors of the Dicke Hamiltonian if the qubit free-term were zero. They have the form:
\begin{equation}
 \left\vert \varphi_{m,k}\left(G\right)\right\rangle=\frac{\left(\hat{A}^{\dag}_m\right)^k}{\sqrt{k!}}\left\vert-mG\right\rangle,
\end{equation}
where $\hat{A}^{\dag}_m=\hat{a}+mG$ and $\left\vert-mG\right\rangle$ is a coherent state that works as a `displaced vacuum', and therefore these states can be seen as displaced Fock states. The basis consists of vectors $\left\vert \Phi _{m,k}\left( G\right) \right\rangle$ where $m$ ranges over all the non-negative eigenvalues of $\hat{J}_x$ and $k$ ranges over all non-negative integers; except when $m=0$, where only even values of $k$ are relevant. The truncation is performed on the displaced Fock states in terms of the values of $k$, from $k=0$ and up to a maximum value $M$. Truncation values as low as $M=8$ are enough to exactly diagonalize the Hamiltonian.
\\
\\
Despite the usefulness of the adapted basis in Eq. \ref{eqDickbasis} for diagonalizing the DM at any value of $\lambda$, when this basis is used to dynamically simulate the adiabatic crossing, the computing time becomes prohibitively large. This is true even for system sizes of just $N = 2^5$, well below the size at which scaling effects are clearly manifested. In this case, dynamical evolution requires a further step. It turns out that the condition of near-adiabaticity is a great advantage, and the strategy is to solve the dynamics in terms of instantaneous eigenstates, as in Eq. 4 of the main text, focusing on the lower part of the spectrum. It requires being careful in continuously interpolating each eigenstate $\left\vert \varphi_{\lambda(t)} ^{(n)}  \right\rangle$ as $\lambda$ varies, even in the presence of (i) level crossings, (ii) almost critical behavior around the phase boundary, or (iii) spurious phases at each instance of diagonalization. This strategy was confronted with dynamical solutions for systems small enough to be directly evolved with the adapted basis, yielding excellent matching. It was even employed in the case of the LMGM and TFIM, in order to set more confidence on its equivalence, and also to ensure the accuracy of those results.

\section{Extensions to non-linear annealing}

In the main text, results were presented for an annealing parameter with linear time-dependence, $\lambda(t) = \upsilon t$. However, we stress that any power-law dependence of the form $\lambda(t) = \upsilon \mathrm{sign}(t) |t|^{\kappa}$ will exhibit this same size-independent behavior. The generalization to this case would imply the phase term in Eq. 4 of main text becoming

$$\exp \left( i \frac{\upsilon^{-1/\kappa}}{\kappa} \int_0^{\lambda} \mathrm{sign}(\lambda \prime) |\lambda \prime|^{\frac{1-\kappa}{\kappa}} \Delta_{n,m}^{(N)}  (\lambda \prime)\mathrm{d}\lambda\prime \right)\ \ .$$

\noindent At the same time in Eq. 7 of the main text, the phase would be redefined as $\Phi_{n,m} (x) = \int_0 ^x \mathrm{sign}(x \prime) |x\prime|^{\frac{1-\kappa}{\kappa}} D_{n,m}(x \prime) \mathrm{d} x \prime$, while $\mu = \frac{\kappa \nu}{\kappa \nu z + 1}$ and $\Lambda = \kappa^{-\mu} \upsilon^{\mu / \kappa} N $. It is clear that the case $\kappa = 1$ would imply less convoluted formulas and that is why it was chosen as the example in the main paper.

\begin{figure}[t]
  \includegraphics[width=0.99\textwidth]{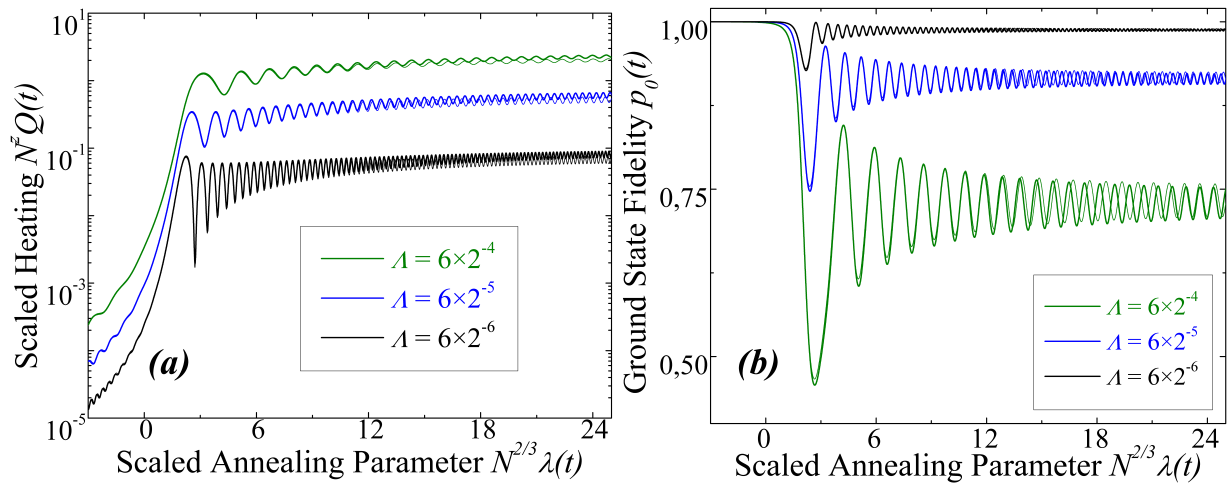}
  \caption{Extended data interval corresponding to results shown for the DM in Fig. 3-c of main manuscript. (a) Scaled heating. (b) Ground state fidelity.
  Data for $N=2^8$ are plotted as thin curves while data for $N=2^9$ are plotted as thick curves.}\label{figsupp}
  \end{figure}

\section{Collapse well beyond the critical threshold}

It might be asked whether some notorious differences between system sizes $N$ in panels of Fig. 1 should be present in the respective panel of Fig. 3 if bigger values of $\lambda$ of the main manuscript were investigated. In order to better appreciate that collapsed curves in Fig. 3 are effectively the continuous dynamical scaled version of data in Fig. 1, a larger range of values of $\lambda$ is illustrated in SM-Fig.\ref{figsupp} for the DM. The chosen interval goes up to $\lambda = 0.62$ for $N=2^8$ and to $\lambda=0.39$ for $N=2^9$ in the $\lambda>0$ ordered phase present in Fig. 1. Therefore, these new plots have a time scale comparable to that of Fig. 1 of main manuscript. In order to have clear curves with a fine mesh resolution, thinner curves have been used for $N=2^8$ instead of symbols. SM-Fig.\ref{figsupp} shows that most differences between system sizes $N$ in Fig. 1 of main manuscript are either because continuous finite-size scaling has not yet been applied and/or slightly because mesh resolution is not enough to show the complete oscillations appreciable in Fig. 3 of main manuscript. In SM-Fig.\ref{figsupp}, it is clear that collapse in both heating and fidelity is preserved well beyond the critical point albeit two non relevant discrepancies emerge for significatively large values of $\lambda$. The first one occurs only for the heating, and it consists of a gradual departure between analogous curves with different $N$. This departure is consistent with the fact that, at very high values of $\lambda$, the scaling of heating by $N^z$ is no longer necessary and a collapse in terms of non-scaled heating, as seen in Fig. 1 of main manuscript, is what it must be expected. The second difference is an occasional loss of oscillation synchronization in the curves corresponding to different $N$. This discrepancy is also negligible, as the mean value of fidelity curves collapses very thoroughly. Both discrepancies are in fact connected, since the cause of oscillations is the accumulated phase difference between instantaneous eigenstates in terms of their respective energy difference, hence the loss of synchronization is also due to the fact that energy differences no longer scale with $N^z$ for high values of $\lambda$.

\bibliography{bibliographySupp}